# Alexander's Wholeness as the Scientific Foundation of Sustainable Urban Design and Planning

Bin Jiang

Faculty of Engineering and Sustainable Development, Division of GIScience
University of Gävle, SE-801 76 Gävle, Sweden
Email: bin.jiang@hig.se

*"... there is such a thing as living structure, and there is an objective distinction between systems which have relatively more living structure and systems which have relatively less living structure."*

Christopher Alexander (2002–2005, Book 2, p. xvi)

**Abstract**
As Christopher Alexander conceived and defined through his life's work – *The Nature of Order* – wholeness is a recursive structure that recurs in space and matter and is reflected in human minds and cognition. Based on the definition of wholeness, a mathematical model of wholeness, together with its topological representation, has been developed, and it is able to address not only why a structure is beautiful, but also how much beauty the structure has. Given the circumstance, this paper is attempted to argue for the wholeness as the scientific foundation of sustainable urban design and planning, with the help of the mathematical model and topological representation. We start by introducing the wholeness as a mathematical structure of physical space that pervasively exists in our surroundings, along with two fundamental laws – scaling law and Tobler's law – that underlie the 15 properties for characterizing and making living structures. We argue that urban design and planning can be considered to be wholeness-extending processes, guided by two design principles of differentiation and adaptation, to transform a space – in a piecemeal fashion – into a living or more living structure. We further discuss several other urban design theories and how they can be justified by and placed within the theory of wholeness. With the wholeness as the scientific foundation, urban design can turn into a rigorous science with creation of living structures as the primary aim.

## 1. Introduction
Modernist urban design, as well as modernist architecture, has been repeatedly criticized for lacking a robust scientific underpinning of its own. City rebuilding and planning are based on *"a foundation of nonsense"* (Jacobs 1961, p. 13), and their theories have been accused of being pseudoscientific (e.g. Cuthbert 2007, Marshall 2012). This situation constitutes one of the motivations of the present paper. However, there is a major difference between the 1960s and the 21st century, when a new scientific foundation for architecture and design was laid out by Alexander (2002–2005), although the mainstream architecture and urban design has yet to adopt the fundamental design thinking. This new scientific foundation is built on the concept of wholeness. The wholeness differs from the concept of wholeness as conceived in biology (Goldstein 1939, 1995), in Gestalt psychology (Köhler 1947), and in a variety of religious and therapeutic contexts, because it is mathematically defined as a recursive structure, recurring physically in space and matter, and reflecting psychologically in human minds and cognition (Alexander 2002–2005; c.f. Section 2 for more details). In this connection, it is very much like wholeness in quantum physics as defined by Bohm (1980).

The wholeness was conceived, developed, and further refined during the 27 years between 1975 and 2002, and finally presented through his magnum opus entitled *The Nature of Order*, a four-volume book focusing on the art of building and the nature of the universe (Alexander 2002–2005). Although wholeness exists pervasively in our surroundings – artifacts, ornaments, rooms, buildings, gardens, streets, cities, and even the entire Earth's surface – it is rather obscure, not so obvious to be seen. Because of this obscure nature of wholeness, Alexander (2002–2005) used hundreds of pictures, paintings, and drawings to illustrate the deep structure of wholeness. As he admitted, there was no mathematics at that time that was powerful enough to capture the wholeness of space. A mathematical model of wholeness has since been developed that it is able to address not only why a structure is beautiful, but also how much beauty the structure is (Jiang 2015). This situation constitutes another major motivation and catalyst of the present paper. The mathematical model is based on a topological representation of wholeness e.g., topology of streets as well as that of cities (Jiang 2018). The topology refers to the topological relationship among geometrically coherent entities rather than geometric primitives of points, lines, polygons, and pixels.

Under the new circumstance, this paper is intended to advocate the wholeness as the scientific foundation of sustainable urban design and planning. The wholeness or living structure, which can be pervasively seen in traditional cities or traditional societies in general (Alexander 2002–2005, Hakim 2009), should become a shared notion of quality of what architects and urban designers are collectively aiming for. With the living structure as the shared notion of quality, the goodness of buildings and cities is considered to be a matter of fact – rather than opinions or personal preferences as currently conceived – and the successes and deficiencies of buildings and cities can be judged objectively or structurally. Although this sounds radical, it is indeed what Alexander (2002–2005) has been advocating and fighting for throughout his entire career (Alexander et al. 2012).

The contribution of this paper is three-fold. First, our argument is based on the newly developed mathematical model of wholeness and its topological representation. Through the model, the concept of wholeness – or Alexander's design philosophy – becomes more accessible thus less obscure to urban design and planning professionals, as well as to other researchers. Second, we argue that living structure is governed by two fundamental laws – scaling law and Tobler's law – that are also what underlies the 15 structural properties (Appendix A), which further serve as transformational properties to transform a space to a living structure. Third, we illustrate that urban design and planning are considered to be wholeness-extending processes through two design principles of differentiation and adaptation, and further discuss on how other urban design theories can be justified and placed within the theory of wholeness, thus wholeness being really a larger theory.

The remainder of this paper is structured as follows. Section 2 introduces the concept of wholeness and two fundamental laws – scaling law and Tobler's law – that govern the deep structure of wholeness. Section 3 presents the mathematical model of wholeness, which is well supported by the topological representation of wholeness, capturing two spatial properties: "far more smalls than larges" (scaling law or spatial heterogeneity) on the one hand, and "more or less similar" (Tobler's law or spatial dependence) on the other. By focusing on how to generate a living structure for a design purpose, Section 4 discusses two design principles – differentiation and adaptation – that are in line with the two fundamental laws of wholeness. Section 5 examines how classic urban theories can be placed within the theory of wholeness in order to further argue why the wholeness can serve as the scientific foundation of sustainable urban design and planning. Finally, Section 6 concludes the paper and points to future work.

**2. The wholeness and its two fundamental laws: Scaling law and Tobler's law**
The wholeness is defined as a recursive structure that recurs in space at different levels of scale; the wholeness is *"a real structure"*, *"nearly a substance"*, not just *"a general appreciation for the unity"* (Alexander 2002–2005, Volume 4, p. 319). The wholeness is made of "far more low-intensity centers than high-intensity ones". The centers are geometrically coherent entities in that space, and they can only be defined by other centers – centers are made of other centers reflexively or recursively. This is somehow like the situation of fax machines. A single fax machine is useless and it has to rely on the other fax machines for its usefulness; the more other fax machines there are, the more useful the fax machine is. All centers at different levels of scale tend to nest or overlap each other to form a complex whole. The intensity of the centers in that space is actually their degree of coherence or life or beauty, which is another name of wholeness (Alexander 2002–2005). In other words, life or beauty or coherence is a quality of space – *"what we perceive as the quality of buildings and artifacts"* (Alexander 2002–2005, Volume 1, p. 110) – that comes about because of the wholeness. A structure with a high degree of wholeness is called a living structure, while a structure with a low degree of wholeness is called a dead or nonliving structure.

Space is neither lifeless nor neutral, but is a living structure – like a growing plant – that is capable of being more living or less living. This new view of space, as conceived by Alexander (2002–2005), represents a paradigm shift from Newtonian and Leibnizian views of space, which are framed under the mechanistic world picture of Descartes (1637, 1954). In contrast, the new view of space is framed under a modified picture of the universe – Alexandrine organic world picture. Under the organic world picture, Alexander (2002–2005, Volume 1, p. 4) made two bold claims: (1) *"all space and matter, organic or inorganic, has some degree of life in it, and that matter/space is more alive or less alive according to its structure and arrangement"*, and (2) *"all matter/space has some degree of "self" in it, and that this self, or anyway some aspect of the personal, is something which infuses all matter/space."* Either of these two claims would radically change our picture of the universe from the mechanistical one to the personal one.

The mechanistic world view makes good architecture virtually impossible because of the lack of a universal standard about the goodness of architecture. As Alexander remarked (2002–2005, Volume 1, p. 16), "*The picture of the world we have from physics, because it is built only out of mental machines, no longer has any definite feeling of value in it; value has become sidelined as a matter of opinion, not intrinsic to the nature of the world at all*". A good building is therefore conventionally considered to be an opinion or personal preferences. This view about goodness of building is very bad for urban design and architecture, because under the view it is virtually impossible to make great buildings or cities. In this respect, the wholeness or living structure provides a universal standard or a shared criterion for the goodness of buildings, cities, or artifacts.

Beauty or life or coherence is essentially structural, not merely an impression or cognition. To illustrate, Figure 1 illustrates some structures with different degrees of wholeness. For example, compared to the two rectangles, the one with a tiny dot has a higher degree of wholeness than the empty one. This is because the rectangle with a tiny dot is able to induce about 20 centers, among which the "far more low-intensity (or life) centers than high-intensity ones" constitute a complex network (Jiang 2015, p. 479–480, Alexander 2002–2005, Volume 1, p. 81–82). In comparison with the empty rectangle and the ellipse, the reader might prefer the ellipse to the rectangle. However, structurally speaking, or as a matter of fact, the empty rectangle has a high degree of wholeness than the ellipse. Why? The empty rectangle has five centers – the rectangle itself and its four corners – whereas the ellipse is able to induce no other center, except for the ellipse itself as the only center. More importantly, the degree of wholeness of a structure is assessed not only in terms of things within it, but also in terms of things surrounding it. The ellipse is not well shaped in terms of its surrounding, since it induces concave rather than convex space. This is the same for the circle, when compared with the square in the second row. This is the reason why the square column is usually more coherent than the round column (Alexander 2002–2005, Volume 1, p. 129). Having argued that both ellipse and circle are not well shaped, we are not suggesting that ellipses or circles should not be used in buildings.

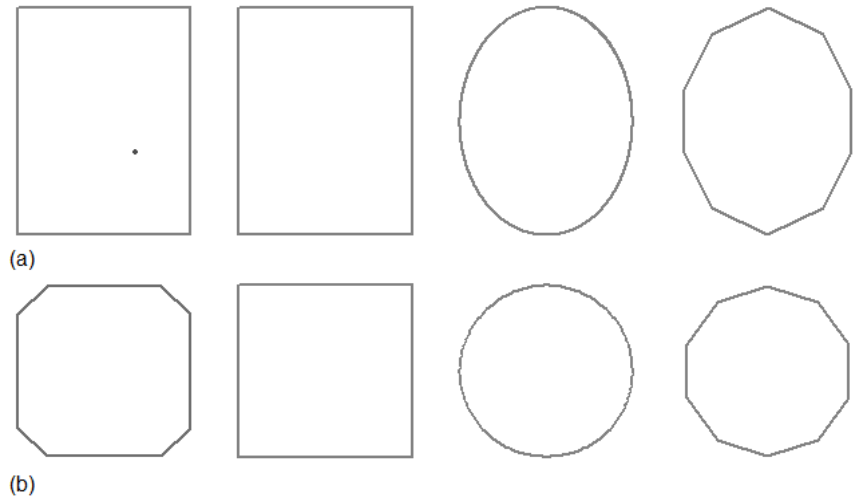

Figure 1: Illustration of physical and mathematical nature of wholeness.
(Note: For the first three columns, the degree of life or wholeness in each row of (a) and (b) decreases from left to right. The reason why the ellipse and the circle have the lowest degree of life concerns their surroundings that are not well-shaped, concave rather than convex. If both the ellipse and circles became polygons while still retaining the overall shapes, the effect will be dramatically different – their degree of life will become far more than the corresponding rectangle and square.)

Instead, our point is that if the ellipse and circle are used, there should be some measure to avoid concave space at their surroundings, so that the part of space in the surroundings still looks convex rather than concave. The octagon has a higher degree of wholeness than the square, because the former is more differentiated than the latter. This is the exact reason why the columns with the octagonal chamfer are more coherent than the square columns (Alexander 2002–2005, Volume 1, p. 113). If the circle and the ellipse are transformed to the decagon and the 10-sided polygon in the fourth column of Figure 1, their degree of wholeness would be improved significantly. This is a very powerful and subtle way of making and intensifying centers, similar to the flutes of the columns in the Temple of Hera at Paestum (Alexander 2002–2005, Volume 1, p. 131–132). As a general rule, more coherent centers tend to lead to a living structure, given that these centers are adapted to each other. For example, the decagon is more alive than the circle because the former tends to induce more centers.

Table 1: Comparison between scaling law and Tobler's law
(Note: These two laws complement each other and recur at different levels of scale in living structure.)

| Scaling law | Tobler's law |
|---|---|
| far more small things than large ones | more or less similar things |
| across all scales | available on one scale |
| without an average scale (Pareto distribution) | with an average scale (Gauss distribution) |
| long tailed | short tailed |
| interdependence or spatial heterogeneity | spatial dependence or homogeneity |
| disproportion (80/20) | proportion (50/50) |
| complexity | simplicity |
| non-equilibrium | equilibrium |

A living structure possesses many, if not all, of the structural properties (Appendix A), so the degree of wholeness of a structure can be judged in terms of how many properties it has. Usually, the more these structural properties, the higher degree of wholeness, meaning more alive or more beautiful or more coherent something is. In many cases, one can simply count the number of local symmetries to determine which of the two structures has a higher degree of wholeness. Underlying the 15 properties are two fundamental laws – scaling law and Tobler's law (1970) – that well complement each other (Table 1, Jiang 2018b) for better understanding and accounting for living structure in our surroundings. Scaling law states that there are "far more small centers than large ones" across all scales ranging from the smallest to the largest. The very first property, *"levels of scale"*, essentially reflects the scaling law. For example, there are "far more low buildings than high ones", and further there are "far more short streets than long ones" in it. Salingaros and West (1999) formulated a universal rule for the distribution of sizes. This universal rule is essentially based on power laws, but scaling law could imply power laws, lognormal, exponential functions or even skewed normal distributions. Tobler's law is widely known as the first law of geography, indicating that nearby centers tend to be "more or less similar". Many of the 15 properties reflect Tobler's law, such as, alternating repetition, local symmetries (at a same scale rather than different scales), deep interlock and ambiguity, contrast, gradients, and simplicity and inner calm; see Appendix A for more details. The 15 properties are often shown in various living structures, yet they should not be followed literally, and they are an outcome rather than a recipe.

The wholeness is not only physical, but also psychological, reflected in our minds and cognition. Due to this cognitive nature of wholeness, we can rely on the human observer as a measuring instrument to objectively judge the degree of life or beauty. Two structures are put side by side and the subject is requested to pick the one that better represents himself. This is the so-called mirror-of-the-self experiment (Alexander 2002–2005, Volume 1, p. 314–350, Wu 2015, Rofé 2016). Although engaging with the human observer, the purpose of the mirror-of-the-self experiment is *not* to seek inter-subjective agreements like ordinary psychological or cognitive tests, but to seek the objective existence of the wholeness. Guided by the 15 properties or tested through the mirror-of-the-self experiment, we can objectively –yet relatively just as we evaluate temperature for warmness – judge the degree of beauty of a pair of structures.

The feeling of wholeness is objective, thus invariant among people across different faiths, ethics, and cultures; *"Ninety percent of our feelings is stuff in which we are all the same and we feel the same things"* (Alexander 2002–2005, Volume 1, p. 4). This ninety percentage should not be understood literally, but rather metaphorically, indicating a majority rather than a minority. In addition to the mirror-of-the-self experiment, we can also determine the degree of wholeness, through the mathematical model of wholeness (Jiang 2015). Through the model, we are able to give an accurate account of why a structure is beautiful and how beautiful it is. This is achieved firstly through a topological representation of the wholeness (e.g. Jiang 2018a), and then by using Google's PageRank scores (Page and Brin 1998) and the ht-index (Jiang and Yin 2014) to characterize the beauty of individual centers and the whole. The topological representation of cities as a coherent whole (Jiang 2018a) reflects both scaling law – at the global scale or across all scales – and Tobler's law at each of these scales.

### 3. A mathematical model of wholeness
The wholeness is made of recursively defined centers, which further constitute a topological structure of relationship of these centers. Based on the topological representation of numerous nested and overlapped centers and their relationships, the mathematical model of wholeness is able to compute the degree of life or beauty for individual coherent centers, as well as for that of the whole. To illustrate, we use a simple mandala as a working example (Figure 2). As shown in Figure 2a, there are many centers – more precisely, "far more small centers than large ones" – and, importantly, small centers are embedded in large ones. We can see many of the 15 structural properties, such as *"levels of scale"*, *"strong centers"*, and *"thick boundaries"*, so there is little doubt that this is a living structure. Figure 2b shows the topological representation of these centers. Note that the node locations of the topological representation in Figure 2b have no effect on computing the degree of wholeness.

The topological representation of these centers is a de facto complex network (Jiang 2015), on which Google's PageRank scores (Page and Brin 1998) are used to indicate the degree of wholeness for individual centers. The degree of wholeness or life or beauty is represented by node sizes: the larger the nodes, the higher the degree. For all of the centers as a whole, there are obviously "far more low-intensity centers than high-intensity ones". The extent to which there are "far more smalls than larges" indicates the degree of wholeness of the whole, and it is measured by the ht-index, a head/tail-breaks-induced index for measuring the complexity or scaling hierarchy of a structure or pattern (Jiang and Yin 2014). The degree of beauty for the mandala as a whole is 5, indicating that the scaling of "far more low-intensity centers than high-intensity ones" recurs four times.

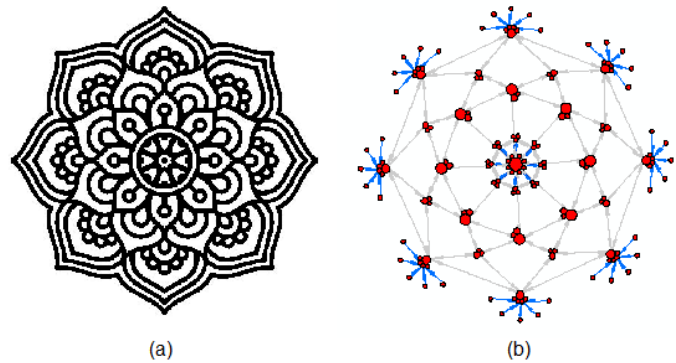

Figure 2: (Color online) A simple mandala (a) and its topological structure of wholeness (b) (Note: The structure of wholeness is essentially topological (b) and it captures the overall character of the mandala, whereas the mandala has geometric details of locations, sizes, shapes, and directions (a). With the topological representation to the right, the node locations have no meaning at all, and only relationship matters. There are two kinds of relationship, indicated by two colors: gray for an undirected relationship and blue for a directed relationship. The node sizes indicate the degree of wholeness or life or beauty or coherence.)

The topological representation of wholeness is a graph theoretic representation of centers, rather than an ordinary graph representation among geometric primitives of points, lines, polygons, and pixels (Bian 2007, Longley et al. 2015), for these geometric primitives are not coherent wholes or centers. Current representations of space in mathematics and physics and geography (including geographic information science as well) reflect essentially local scales; *"It is precisely this assumption about space which is being challenged by the idea of the field of centers"* (Alexander 2002–2005, Volume 1, p. 460). In terms of Alexander (2002–2005, Volume 1, p. 458–459), *"The extraordinary thing does not lie in the 'system' of these centers, not in the fact that they cooperate to form a system. ... What is extraordinary here is something else. ... The life of any given center depends on the whole field of centers in which this center exists. This means that the most fundamental property of each center – its degree of life – is defined not by the center itself but by its position in the entire field of centers. ... This is the thing which is peculiar. It is a type of behavior which is not typical of Newtonian space at all. Indeed, it is a type of behavior which is also not typical of relative space, nor even of quantum mechanical space. ... Thus the intensity of a center can never be understood as a local property of that center itself, merely in terms of its own local structure. ... This is the essence of the recursive definition of a center."*

It is not easy to identify centers from a whole and they must be in line with our cognition. In a city, street segments are not centers since they are not congruent with our cognition, but individual streets generated from the street segments can be considered to be coherent centers. In a city image, individual pixels are not centers, but individual meaningful places are centers, since they tend to meet both scaling law and Tobler's law. There are "far more small cities than large ones" recurring at the country scale, so all cities in a country constitute a coherent whole (Jiang 2018a). This topological representation of cities can be considered to be the randomized or statistical model of central place theory (CPT) (Christaller 1933, 1961); see Section 5 and Figure 5 for further details. Just like a city or country, the entire Earth's surface is a living structure, since there are "far more small countries than large ones" recurring at the global scale.

The mathematical model of wholeness implies a new mathematical structure of space, in which the intensity or life of a center is a global effect rather than a local effect. The global effect implies that every center can – through the underlying structure of wholeness – affect other centers that are not nearby. As an example, the living structure or wholeness of the Alhambra plan is a field of centers (Figure 3), in which the intensity is a global effect. Changing one local part would likely affect all other centers in the space, and even beyond towards a larger whole that contains the space. This global effect reflects in both the topological representation and the way in which the intensity or life is calculated.

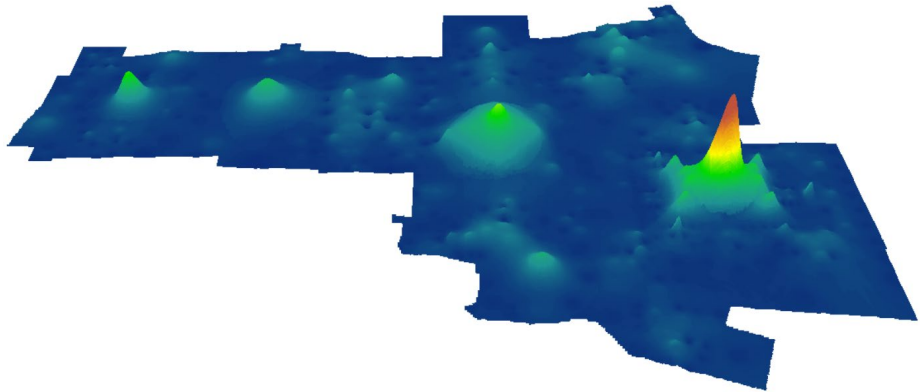

Figure 3: (Color online) The field of centers of the Alhambra plan
(Note: The field or surface of centers was interpolated from the degrees of life of the individual centers, calculated from the mathematical model of wholeness (Jiang 2015). The spectral colors show the degree of life with red for the highest, blue for the lowest, and other colors for those between the highest and the lowest.)

## 4. Wholeness-extending process: Differentiation and adaptation

Having discussed the physical and mathematical nature of the wholeness, as well as its reflections in our minds and hearts, urban design is essentially concerned with how to create the kind of structure in our built environment. When they are used to generate living structure or wholeness, the 15 properties (Appendix A) become transformation properties. All the 15 transformation properties can be summarized by two design principles – differentiation and adaptation – through which a space or structure is recursively differentiated to induce many nested and overlapped centers that are well adapted to each other to be a coherent whole. In Figure 1, the tiny dot is able to differentiate that paper, and induce about 20 centers (Alexander 2002–2005), and eventually to create a more coherent whole than that of a blank paper. In the same way, the decagon and the 10-sided ellipse shape are able to create more centers than their counterparts of the circle and the ellipse. Therefore, this process is also called the centering process, or structure-preserving process. Throughout the four volumes, in particular Volume 2 entitled *"The Process of Creating Life"* and Volume 3 entitled *"A Vision of a Living World"*, Alexander (2002–2005) provided hundreds of practical examples on how to generate– in a piecemeal fashion – living structures at different levels of scale of built environment. In particular, the column example provides a step-by-step recursive process for transforming a simple cylinder into a living column (Alexander 2002–2005, Volume 1, p. 128–130, Figure 4).

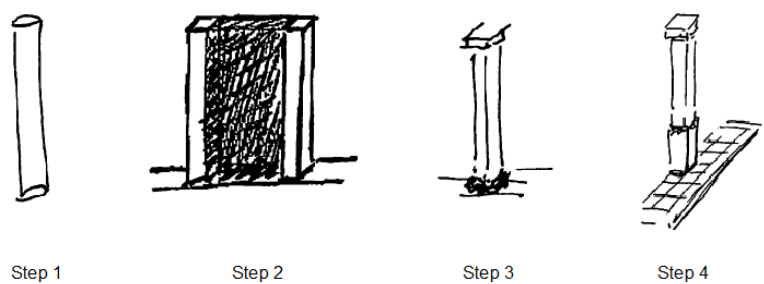

Figure 4: A living column is transformed – in a step by step recursive fashion – from a cylinder

Let us use 10 numbers that follow exactly Zipf's law (1949): 1, 1/2, 1/3, …, and 1/10 as a working example to further illustrate two design principles or underlying the wholeness-extending transformations. The 10 numbers meet the scaling law statistically or the transformation property of *"levels of scale"*. Why? The mean of the 10 numbers is 0.29, which partitions the 10 numbers into two parts: the first three, as the head (accounting for 30%), are greater than the mean, and the remaining seven (70%) as the tail are less than the mean. For the three numbers in the head, their mean is 0.61. This mean further partitions the three numbers into two parts: 1 (33%) as the head is greater than the mean 0.61, and 1/2, and 1/3 (67%) as the tail are less than the mean. As seen from the above head/tail breaks process (See below the recursive function, Jiang 2013a), the notion of "far more smalls than larges" recurs twice, so there are three hierarchical levels among the 10 numbers: [1], [1/2, 1/3], [1/4, 1/5, …, 1/10].

Recursive function of head/tail breaks

```
Function Head/tail Breaks:
     Rank the input data from the largest to the smallest
     Break the data around the mean into the head and the tail;
    // the head for those above the mean
    // the tail for those below the mean
    While (length(head)/length(data)<=40%):
        Head/tail Breaks (head);
End Function
```

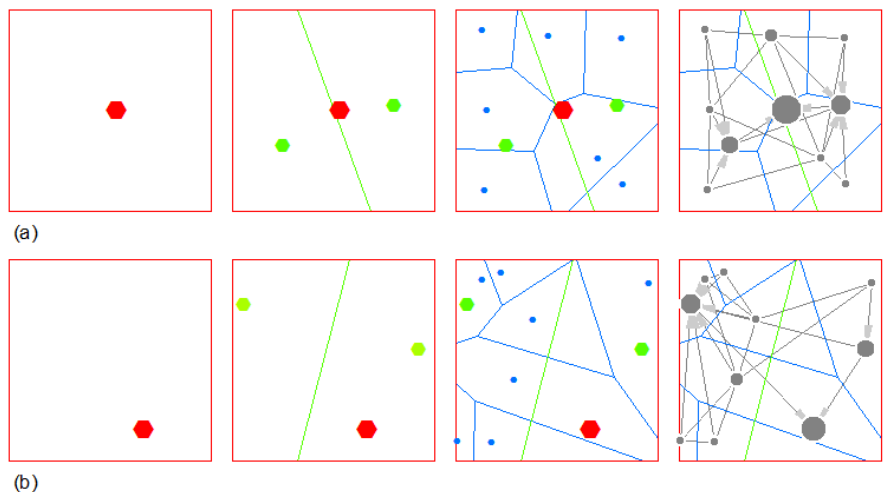

Figure 5: (Color online) Unfolding of two different configurations
(Note: There are 10 cities with sizes that follow Zipf's law exactly: 1, 1/2, 1/3, …, and 1/10. The first three panels from the left indicate how the two structures are unfolded or transformed. Statistically, both structures are living structures, since they contain "far more smalls than larges". However, geometrically – or topologically, to be more precise – the pattern of the second row (b) is less coherent or less beautiful than that of the first row (a). This fact can be seen from several aspects: (1) the largest red dot in row (b) is far from the center, so it tends to detract rather than to reinforce the whole; (2) the two green dots are not well located as these two in row (a), so detract rather than reinforce the two green sub-wholes; (3) at the lowest hierarchy, or the blue Thiessen polygons in row (b), look too diverse rather than "more or less similar" as in row (a), so violating Tobler's law; and (4) the two lowest hierarchical levels are not well adapted: six blue dots under the left green dot's polygon, while only one blue dot under the right green dot's polygon.)

What we wanted to do is to locate these ten numbers – in a step by step fashion – in a square space to constitute a coherent whole. As shown in Figure 5 (first row a), we first put the largest number in the middle of square, and then two middle ones near the largest, and finally the remaining seven smallest. This is one of the coherent wholes for the 10 numbers in the space using the two design principles. On the one hand, the square is recursively differentiated: first two parts by the green line, and later seven parts by blue lines, with "far more smalls than larges" across the three scales. On the other hand, the induced centers, or the partitioned parts, are well adapted to each other with "more or less similar" at each scale, thus meeting Tobler's law. In comparison, we created another pattern shown in Figure 4 (second row b). The scaling law remains unchanged for it is determined by the same 10 numbers.

However, the induced centers of the second pattern at the smallest scale are not well adapted to each other, thus violating Tobler's law to some extent, as can be seen from the partitions by the blue lines. More importantly, across the middle and smallest scales, centers are not well adapted to each other; the left green contains six blue points, whereas the right green contains only one blue point. Therefore, the second pattern – by violating Tobler's law – is apparently less coherent or less beautiful or with less life than the first. This coherence difference between the two patterns or structures is partially reflected in the calculated degrees of beauty shown in the last column of Figure 5, where the largest dot in row a is apparently a bit larger than the largest dot in row b.

What was just discussed is, of course, a very simple example, which relies on topological representation for illustrating different configurations. Actual architectural and urban design is far more complicated than this example, but the underlying design principles of differentiation and adaptation remain the same; more specifically, the 15 transformation properties (Appendix A) remain the powerful means to reach a living structure. The wholeness-extending process is the only way to create living structures. Based on the wholeness theory, urban design should be considered to be a piecemeal and iterative growth process; every transformation is based on what exists, identifying latent centers and further extending or enhancing these centers to be more living.

Relying on human feeling or the feeling of wholeness, right decisions, at each iteration, can be made to ensure that the degree of wholeness is increased rather than decreased. With the help of the mathematical model of wholeness, computer algorithms can be developed to guide the generation of living structures. Urban design is the process of continuously generating living centers: "far more small centers than large ones", with "more or less similar centers" nearby through differentiation and adaptation. This wholeness based urban design *"is different, entirely, from the one known today"*, and *"the task of creating wholeness in the city can only be dealt with as a process"* (Alexander 1987, p. 3). The 15 transformation properties (Appendix A) provide not only clues or hints, but also effective means for urban design processes.

Many efforts have been made to advocate the wholeness-oriented transformations for sustainable urban design. Salingaros (2012, 2013) developed 12 lectures on architecture for algorithmic sustainable design, and a unified architectural theory for transforming our built environment for the better through adaptive and sustainable design. Mehaffy and Salingaros (2015) explored many complexity science ideas, such as fractals and complex networks, which have close links to the wholeness for designing a living planet. Importantly, fractal or living structure is found to have a healing effect for human health and well-being (Ulrich 1984, Taylor 2006, Salingaros 2012b). The wholeness constitutes some key conceptual insights for the new urban renaissance for making or re-making better built environments (Mehaffy 2017, Salingaros 2019). The wholeness-oriented making or design is not just a vision, but a mission that many of urban designers are aiming for. Based on the legacy of Alexander and his life-long pursuit of beauty (Salingaros 2018), the Building Beauty program (https://buildingbeauty.org/) has started this mission in educating the next generation of architects and urban design professionals. All of these efforts will hopefully transform urban design towards a well-respected discipline – design science.

**5. The wholeness accounting for other urban design related theories**
The wholeness serves as the scientific foundation for sustainable urban design and planning because it reveals not only the nature of beauty: being objective and structural, but also the nature of space: neither lifeless nor neutral, but a living structure. More importantly, it can be used to account for many other urban design and planning theories such as the image of the city (Lynch 1960), space syntax (Hillier and Hanson 1984), the city as an organized complexity (Jacobs 1961), and the CPT (Christaller 1933, 1966). Strictly speaking, none of these theories really focuses on urban design and planning. Instead, they concentrate largely on understanding the complexity of cities, rather than on making or re-making cities. This is the same for fractal geometry. Fractal geometry is mainly for understanding complexity of fractals rather than for making fractals, even with limited making, often resulting in fractals of *"pretty, yet pretty useless"* (Mandelbrot 1983); fractal geometry is unable to quantify goodness of fractals. In contrast, living geometry – another name of the wholeness theory – aims not only for understanding complexity, but also for actively creating complex or living structure; living geometry *"follows the rules, constraints, and contingent conditions that are, inevitably, encountered in the real world"* (Alexander et al. 2012, p. 395). This section will further discuss on wholeness and its relationship to the other urban design related theories.

Why the image of the city can be formed in our minds and hearts has little to do with who we are or what gender or ethnic background we have; it is primarily to do with the underlying living structure of "far more smalls than larges". The largest in the top head constitutes the image of the city (Jiang 2013b). The largest refers to not only largest sizes, but also the most connected, or the most meaningful. In other words, the living structure of "far more smalls than larges" can be in terms of geometric sizes, topological relationships, or semantic meaning. For example, a place that does not have the highest buildings or the most connected streets constitutes part of the mental image of the city because it bears the most meaning for a person. Given this new interpretation, it is not so much about geometric or visual shapes, so called the city elements including nodes, paths, edges, districts, and landmarks (Lynch 1960). Among the five city elements, only landmarks make good sense for the image of the city, and other four are irrelevant (Jiang 2013b). In this connection, I would like to refer to this statement again: *"Ninety percent of our feelings is stuff in which we are all the same and we feel the same things"* (Alexander 2002–2005, Volume 1, p. 4). While individuals do have their own image of the city, but that accounts for only ten percent of feelings. Recently, Shushan et al. (2016) bring further empirical evidence for what we discuss here by using virtual environments to verify the image of the city in both heterogeneous and homogenous environments, which represent respectively living and dead structures.

Space syntax is a theory developed by Hillier and Hanson (1984) for urban morphological analysis using graph theoretic representations. The fundamental idea of space syntax is that a space is too big or too complex to be well perceived entirely, so it is partitioned into many small-scale spaces that can be perceived entirely or from a single vantage point of view. All the small-scale spaces constitute an interconnected whole represented by a graph. There are different representations of space, which can be put into two major categories: geometric and topological. The major difference between the two representations lies on whether they capture the underlying living structure of "far more small centers than large ones": topological representation can, while geometric representation cannot. The topological representation is among meaningful geographic features such as streets and cities (Jiang 2018a, 2018b), while the geometric representation is among meaningless geometric primitives of points, lines, or polygons. For example, a city can be topologically represented as an interconnected whole consisting of "far more less-connected streets than well-connected ones". A city can also be topologically represented by axial lines – a so-called axial map – which is not as good as the streets in terms of capturing the underlying living structure. From the perspective of wholeness, a spatial representation that captures the underlying wholeness tends to have high prediction rate for human activities or urban traffic. Geometric representations are unable to capture the living structure because they concentrate on geometric primitives of points, lines, or polygons, each of which makes little sense in our minds and hearts.

A city is essentially the problem of organized complexity, as Jacobs (1961) claimed in the classic work *The Death and Life of Great American Cities*. This complexity science perspective on cities helps better understand not only how a city looks, but also how a city works. For example, a city is not a tree (Alexander 1965), but a semilattice, which exhibits scaling hierarchy of "far more smalls than larges"; a city is self-organized with emergences developed from the bottom up. The understanding of cities from the complexity science perspectives helps in planning and designing cities to become living structures. However, the generative science, including *A New of Kind of Science* (Wolfram 2003), is largely limited to the understanding of complexity, and hardly touches the fundamental issue of making or design of living structure (Simon 1962). Critically, science as it is currently conceived lacks a standard or criterion about goodness of structure, being judged as opinions and individual preferences; therefore, it is no wonder that urban design – or geography in general – has always been a minor science. Urban designers and researchers try to apply ideas of major sciences such as physics, biology, and computer science in order to make their work seemingly "scientific". If the wholeness were adopted as the scientific foundation, and if space were properly understood as a living structure, urban design and

planning – or geography in general – would *"play a revolutionary role in the way we see the world … and will perhaps play the role for the world view of the 21st and 22nd centuries, that physics has played in shaping the world view of the 19th and 20th"* (Alexander 1983, cited in Grabow 1983, p. xi); see Volume 4 of Alexander (2002–2005) for detailed arguments on the modified world view that is organic in essence.

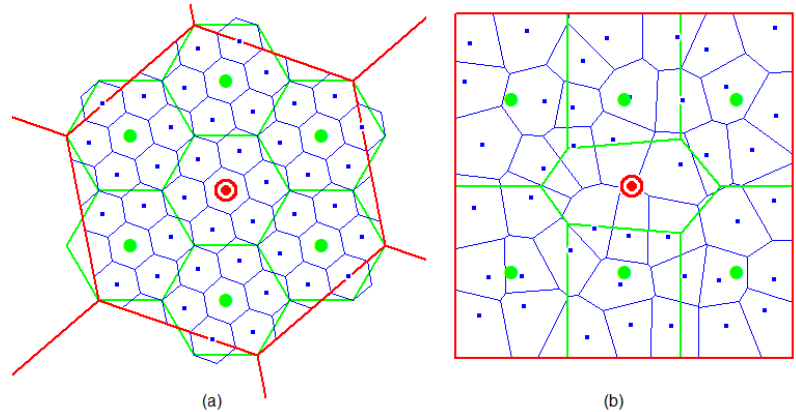

Figure 6: (Color online) A diagram of the classic CPT (a) and its statistical counterpart (b) (Note: The major difference between the classic CPT (a) and its statistical counterpart (b) is that in the former, every city is supported by exactly other six cities, while in the latter, the number of supporting cities vary from one to another. The diagram to the left is the case of k = 7 (Christaller 1933, 1961), in which there three hierarchical levels indicated by three colors with red being the highest, blue the lowest, and green between the highest and the lowest. The randomized or statistical counterpart to the right (b) is a topological representation of wholeness. Given the set of city locations, the city sizes have three hierarchical levels indicated by three colors, which are derived using head/tail breaks (Jiang 2013a). At each hierarchical level, space is differentiated by individual cities and represented by Thiessen polygons. For more details on the topological model, one can refer to Jiang (2018a).)

CPT is a geographic theory that seeks to account for spatial configuration of cities (or human settlements in general) in terms of their numbers, sizes, and locations in a region or country (Christaller 1933, 1966, Chen and Zhou 2006, Jiang 2018a, 2018b). This spatial configuration is essentially a structure of wholeness of different sized cities at different hierarchical levels, and governed by two fundamental laws of living structure: scaling law and Tobler's law. On the one hand, there are "far more small cities than large ones" across all scales ranging from the smallest to the largest; on the other hand, there are "more or less similar sized cities" on each scale. Both laws complement each other to characterize all the cities as a coherent whole (Jiang 2018a). However, like many other classic fractals such as the Koch (1904) curve and the Sierpinski (1915) carpet, the classic CPT is too restricted – thus less living – to be a useful model in reality; every city is surrounded by six other cities at a low level (Figure 6a). Instead, the topological representation of wholeness provides a realistic model – thus more living – that can be applied directly to cities, and it is a de facto randomized or statistical model of the classic CPT (Figure 6b). In other words, the topological model is to the CPT model what the coastline model – or statistical fractal in general (Mandelbrot 1967, 1983) – is to the Koch curve (Koch 1904). To this point, we can better understand that the two configurations shown in Figure 5 are actually based on the topological model, and why one is more coherent than another from the perspective of CPT.

**6. Conclusion**
Relying on the mathematical model and in particular the topological representation of wholeness, we have argued and demonstrated the mathematical and physical structure of wholeness, and how sustainable urban design can be viewed as the wholeness-extending processes. Wholeness is structural and objective, and pervasively exists in our surroundings, such as artifacts, ornaments, buildings, gardens, streets, and cities; and it can trigger positive feelings – life and beauty – in our minds and hearts. There are two design principles – differentiation and adaptation – that guide the wholeness-extending processes for transforming, in a recursive step-by-step manner, a space into a living or a more living structure. Both differentiation and adaptation are what underlie the 15 transformation properties (Appendix A) that were distilled from traditional and vernacular buildings, as well as from nature. In this sense, the wholeness is what Alexander discovered in nature and in traditional buildings or artifacts, rather than his invention. What also underlie the 15 structural properties (Appendix A) are two fundamental laws – scaling law and Tobler's law – respectively for characterizing living structure with "far more small centers than large ones" across all scales, and "more or less similar centers" on each scale.

As the scientific foundation of urban design, wholeness can be used to effectively justify and account for other urban design theories. The image of the city arises out of the underlying living structure of the city; in other words, a city lacking the living structure would not be able to form the mental image in human minds. Why space syntax works relies on the fact that the topological representation – rather than the geometric representation – is able to capture the underlying living structure of space. A city is essentially the problem of organized complexity (Jacobs 1961) or living structure. The classic CPT model and its variant – the topological representation – can be well justified and accounted for by wholeness. Urban design can, should, and must become part of new complexity science (Alexander 2003), which aims for not only understanding, but also for generating the complex or living structure of our built environments. Our future work will point to the development of computer algorithms for automatically generating living structure for sustainable urban design and planning.

**Note by the Author:**
It should be noted wholeness is believed to be an ancient idea that human beings used – unconsciously or subconsciously – in design for centuries. The discovery of living structure or the formulation of the wholeness by Christopher Alexander is new, yet not so new, beginning in the 1980s. This situation can be compared to fractal thought, which is an ancient design idea used in many traditional designs across many cultures and countries, but fractal geometry was established in the 1970s.

**Appendix A: A brief introduction to the 15 properties**

The 15 properties, also called structural or transformational properties as sketched in Figure A1, are what Alexander first discovered after he developed the pattern language work, because the pattern language failed to create living buildings or cities. He thought the 15 properties must be the fundamental stuff that makes a building or city alive. Later on, he further realized that it is wholeness or living structure, or wholeness-extending process that is what underlies living environments. To make the paper self-contained, this Appendix briefly introduces the 15 properties, and associate them with two fundamental laws: scaling law and Tobler's law.

1. Levels of scale: This first property is essentially scaling law of "far more small centers than large ones". Or alternatively, all centers ranging from the smallest to the largest form a scaling hierarchy, thus meeting scaling law.

2. Strong centers: A strong center is a geometrically coherent unit or sub-whole human beings can easily identify, and it is supported by other surrounding centers. The surrounding centers tend to meet Tobler's law at each level of scale, and scaling law across different levels of scale.

3. Thick boundaries: A center often differs or separates from other centers by thick boundaries. Or a thick boundary defines a coherent center or sub-whole, which meets not only scaling law, but also Tobler's law.

4. Alternating repetition: Alternating repetition of centers tend to be more living than monotonic repetition of centers. These different or alternating centers tend to meet Tobler's law.

5. Positive space: Space can be seen from both figure and ground, each of which should be well shaped, by which it means convex rather than concave. In cities, both buildings and the space between the buildings should be well shaped, somehow like a piece cloud in which both white parts and blue parts are well shaped. For a positive space that can be partitioned into many pieces, both scaling law and Tobler's law apply to these pieces.

6. Good shape: A shape is good, if it consists of good shapes. It is essentially a recursive rule for assessing good shapes. A good shape can have multiple levels of scale, thus meeting scaling law.

7. Local symmetries: Local symmetries differ from global symmetry which is defined at the global level of scale. Local symmetries refer to symmetries at smaller levels of scale. At each small level, centers are "more or less similar" – meeting Tobler's law, whereas at the global level or across scales, there are "far more small centers than large ones" – meeting scaling law.

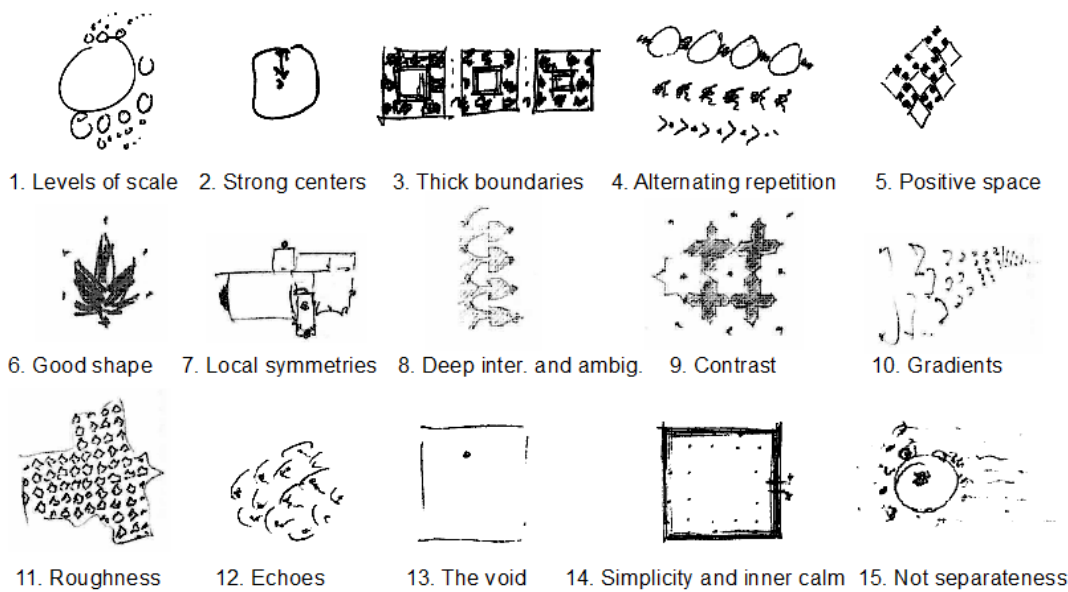

Figure A1: The 15 properties of wholeness (Alexander 2002–2005)

8. Deep interlock and ambiguity: Centers interpenetrate deeply each other, forming deep interlocks, and creating ambiguity among centers. Those interlocked tend to meet Tobler's law.

9. Contrast: Contrast recurs between adjacent centers, which tend to be "more or less similar", thus meeting Tobler's law.

10. Gradients: The gradients property is a field-like character of the center, and it is better characterized by Tobler's law, since nearby centers are "more or less similar".

11. Roughness: Roughness is a key feature of fractals or living structures, yet the degree of roughness tends to meet Tobler's law on a same level, and scaling law across different levels.

12. Echoes: This property of echoes resembles very much to the self-similarity of fractals. It can occur within a sub-whole or across different sub-wholes.

13. The Void: A void is an empty center at the largest scale, surrounded by many small and smaller centers, with which the void meets scaling law.

14. Simplicity and inner calm: This property reflects "more or less similar things" – or even exactly the same things – on a scale within a living center, thus meeting Tobler's law. It should be noted that this property is on a scale rather than across all scales. When applied to all scales, it becomes minimalism.

15. Not separateness: A center is not separable from its surrounding centers. This property has multiple meanings: e.g., centers of a living whole cannot be separated, multiple levels of scale are not separable, a whole is not separable to its larger whole, and the remaining 14 properties are hard to separate from each other. The first and the last properties are probably the most important of all.